# High Spectral-Efficiency, Ultra-low MIMO SDM Transmission over a Field-Deployed Multi-Core OAM Fiber

Junyi Liu[1,†], Zengquan Xu[1,†], Shuqi Mo[1,†], Yuming Huang[1], Yining Huang[1], Zhenhua Li[1], Yuying Guo[1], Lei Shen[2], Shuo Xu[2], Ran Gao[3], Cheng Du[4], Qian Feng[4], Jie Luo[2], Jie Liu[1,5 *], and Siyuan Yu[1]

[1]State Key Laboratory of Optoelectronic Materials and Technologies, School of Electronics and Information Technology, Sun Yat-Sen University, Guangzhou 510006, China.

[2]Yangtze Optical Fibre and Cable Joint Stock Limited Company, State Key Laboratory of Optical Fibre and Cable Manufacture Technology, No.9 Guanggu Avenue, Wuhan, Hubei, China.

[3]School of Information and Electronics, Beijing Institute of Technology, Beijing 100081, China.

[4]Fiberhome Telecommunication Technologies Co. Ltd, Wuhan, 430074, China.

[5]School of Electronics and Information Technology and Guangdong Provincial Key Laboratory of Optoelectronic Information Processing Chips and Systems, Sun Yat-sen University, Guangzhou 510006, China

[†]These authors contributed equally.

*Corresponding author: liujie47@mail.sysu.edu.cn;

## Abstract

Few-mode multi-core fiber (FM-MCF) based Space-Division Multiplexing (SDM) systems possess the potential to maximize the number of multiplexed spatial channels per fiber by harnessing both the space (fiber cores) and mode (optical mode per core) dimensions. However, to date, no SDM transmissions over field-deployed FM-MCFs in realistic outdoor settings have been reported, which contrasts with SDM schemes demonstrated using single-mode multi-core fibers (SM-MCFs) installed in practical fiber cable ducts. In this paper, we present the successful demonstration of bidirectional SDM transmission over a 5-km field-deployed seven ring-core fiber (7-RCF) with a cladding diameter of 178 μm, achieving a Spectral Efficiency (SE) of 2×201.6 bit/s/Hz. This work establishes a new record for the highest SE attained in SDM demonstrations utilizing field-deployed fiber cables, achieving an approximate 10x increase compared to the SE of reported field-deployed optical fiber cable transmission systems. Notably, these results are realized through the utilization of small-scale modular 4×4 multiple-input multiple-output (MIMO) processing with a time-domain equalization (TDE) tap number not exceeding 15, maintaining a complexity per unit capacity comparable to that of MIMO equalization in SDM demonstrations employing weakly coupled SM-MCF cables. These results underscore the significant potential for achieving heightened SE and expanding capacity per individual fiber using SDM techniques in practical applications.

## Introduction

Due to the rapid development of techniques such as wavelength division multiplexing (WDM), advanced modulation formats and coherent detection with digital signal processing (DSP), per-fiber capacity growth of the single-mode fiber (SMF) communication systems has closely tracked the exponential growth of traffic demand over the past few decades[1]. However, this growth trend is becoming unsustainable as the SMF capacity approaches its limits imposed by the fiber nonlinear effects[2], necessitating the development of novel technologies to keep up with traffic demand. In this context, space-division multiplexing (SDM) techniques[3–15], which explore degrees of freedom in the transverse spatial domain of optical fibers to enhance the per-fiber capacity and spectral efficiency (SE) of communication systems, have garnered significant interest in recent years.

Several approaches have been implemented to increase the spatial channel counts and thus the capacity/SE per fiber in the SDM systems, including mode-division multiplexing (MDM) schemes utilizing few-mode fibers (FMFs)[3–6], core-multiplexing schemes employing multi-core fibers (MCFs)[7, 8] and the combination of these two approaches based on few-mode multi-core fibers (FM-MCFs)[9-13,15]. Compared with the former two schemes, the FM-MCF-based SDM systems can potentially provide the highest number of multiplexed spatial channels per fiber by simultaneously utilizing both the optical fiber modes and fiber cores while only require small-scale and, therefore, low-complexity Multiple-Input Multiple-Output (MIMO) processing to handle the intra-core modal coupling, as the inter-core coupling can be disregarded when the core pitch is sufficiently large[6].

Laboratory SDM demonstrations utilizing FM-MCFs have shown significant progress in the past few years, substantially increasing both the per-fiber capacity and SE by orders of magnitude compared with SMF transmission systems[9, 11, 12, 16–20]. However, to the best of our knowledge, no SDM transmissions over field-deployed FM-MCFs in a realistic outdoor environment have been reported. This stands in contrast to SDM demonstrations utilizing single-mode multi-core fibers (SM-MCFs) installed in practical fiber cable ducts, which have been presented in several publications[21–23].

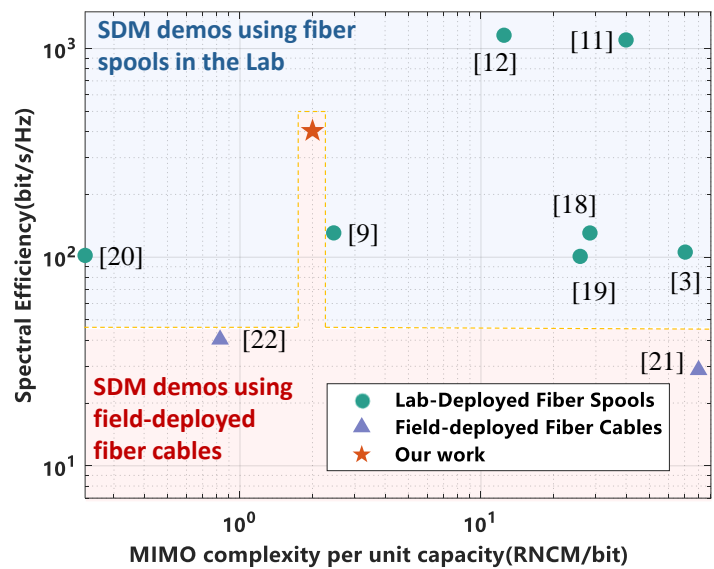

Fig. 1. SE versus MIMO complexity per unit capacity of the SDM demonstrations utilizing lab-deployed FM-MCF spools or field-deployed MCF cables (Data points are evaluated based on information reported from the cited references).

Figure 1 summarizes the SE versus MIMO complexity per unit capacity of recent SDM demonstrations utilizing lab-deployed FM-MCF spools or field-deployed SM-MCF cables. Here the

MIMO complexity per unit capacity is expressed as the required number of complex multiplications (RNCM) per bit [9] (calculation details can be found in Supplement S1). The data indicates that the SE of SDM transmission over field-deployed SM-MCF cables is considerably lower than that demonstrated using lab-deployed FM-MCF spools. This discrepancy primarily arises from the fact that FM-MCFs can provide the highest number of multiplexed spatial channels per fiber by simultaneously utilizing both multiple fiber cores and multiple fiber modes in each core. High-SE transmission over field-deployed FM-MCFs demonstrated in a realistic outdoor environment would take SDM techniques to a new stage towards practical implementations.

In addition to improving SE, several other critical factors must be taken into account for practical applications of SDM transmission over field-deployed FM-MCF cables. Firstly, given the importance of managing power consumption and cost for the transceivers, the complexity of MIMO processing, which addresses inter-SDM-channel crosstalk, should be kept at a minimum level, ideally comparable to that employed in transmission systems over single-mode fibers or weakly coupled SM-MCFs (e.g., MIMO complexity per unit capacity calculated according to information in references [8] shown in Fig. 1). Furthermore, the fiber cladding diameters must be constrained within a specific range to maintain good mechanical stability and the lifetime of optical fibers and fiber cables. Specifically, under the condition of a minimum bend radius of 60 mm and 100 turns, the maximum acceptable fiber cladding diameter to achieve a 1% failure probability over 20 years is 230 μm [24].

In this paper, we present the successful demonstration of SDM transmission over a 5-km field-deployed seven ring-core fiber (7-RCF) with a cladding diameter of 178 μm, achieving a remarkable SE of 403.2 bit/s/Hz. To the best of our knowledge, this sets a record for the highest SE achieved in SDM demonstrations utilizing field-deployed fiber cables. In our demonstration, 84 orbital angular momentum (OAM) mode channels (7 cores × 6 modes × 2 polarizations) are multiplexed, each carrying 40 wavelengths. By implementing bidirectional transmission in the fiber, the number of channels per-fiber can be effectively doubled. Utilizing 12-Gbaud 8-quadrature amplitude modulation (8QAM) per data channel, a raw (net) SE of 2 × 241.92 (2 × 201.6) bit/s/Hz and a capacity of 241.92 (201.6) Tbit/s have been realized. These results are accomplished solely through the application of modular 4×4 MIMO processing with a time-domain equalization (TDE) tap number not exceeding 15, whose complexity per unit capacity is quite comparable to that of the MIMO equalization employed in SDM demonstrations using weakly coupled (SM-MCF) cables, as illustrated in Fig. 1. Furthermore, we also conduct experimental analysis to evaluate the effect of Rayleigh Backscattering (RB) on the bidirectional transmitting signals.

Our demonstration surpasses the SE milestone of 400 bit/s/Hz for SDM transmission systems using field-deployed fiber cables, representing an approximately 10x increase compared to the SE of reported field-deployed optical fiber cable transmission systems. It also simultaneously maintains a low MIMO complexity per unit capacity comparable to that utilized in the SM-MCF based SDM demonstrations, and features a 7-RCF cladding diameter of less than 200 μm. Therefore, these results underscore the considerable potential for achieving higher SE and expanding capacity per single fiber through SDM techniques in practical implementations.

## Methods and Experiment

### 7-RCF cable fabrication and installation

A 5-km fiber optic cable is fabricated with a diameter of 10.7 mm. The cross-sectional views of the fabricated cable are depicted in Fig. 2(a). Twelve optical fibers each with a diameter of 250 μm are wrapped in a jelly-filled loose tube made of polyethylene terephthalate acetate plastic, forming a fiber unit. This fiber unit and other four filling ropes with the same outer diameter are located around the central metal reinforced core. Two layers of polyethylene are wrapped around the exterior, serving as the inner liner and outer protective layer of the fiber cable. The twelve optical fibers in the fiber unit can be categorized into six types, each identifiable by a unique color-coded plastic tape. The specific parameters of the six types of optical fibers are listed in Table 1. Our primary focus in this study is on the 7-RCF incorporated in the cable, employed as the transmission fiber in our experimental demonstration. The cross-section of the 7-RCF is illustrated in Fig. 2(b).

Table 1. Parameters of the optical fibers within the cable

| No. | Color | Diameter ($\mu m$) | Type |
|---|---|---|---|
| 1 | Blue/Orange | 150 | 7-core SMF |
| 2 | Green/Brown | 125 | 1-core SMF |
| 3 | Grey/Red | 125 | 1-core RCF |
| 4 | Black/Yellow | 125 | 1-core RCF |
| 5 | Purple/Pink | 125 | 1-core RCF |
| 6 | Nude/ Black-stripes | 178 | 7-core RCF |

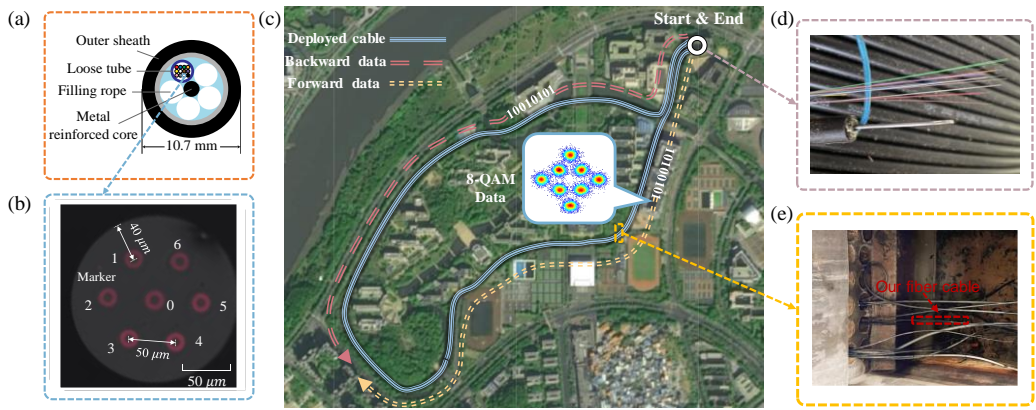

Fig. 2. (a) Cross-section diagram of the fabricated 7-RCF cable; (b) Cross-sectional photo of the fabricated 7-RCF; (c) Schematic diagram of the field-deployed fiber cable route; (d) The optical fibers in the fiber unit; (e) field-deployed fiber cable in a fiber pipeline.

The fabricated fiber cable was deployed in an outdoor environment, and the route is illustrated in Fig. 2(c). The installation commenced at our laboratory, traversing ducts within the State Key Laboratory of Optoelectronic Materials and Technologies (OEMT) building. Subsequently, the cable was laid in the underground cable ducts on the Sun Yat-sen University (SYSU) campus. Following the path of the underground ducts, depicted by the blue line in Fig. 2(c), the cable formed a loop back to the OEMT building, ultimately reaching back to the laboratory. The total length of the installed fiber cable extended approximately 5 km, with ~ 4.5 km within the campus underground pipeline and ~ 0.5 km within the duct of the OEMT building. This configuration allows for experimental setups at both the transmitting and receiving ends to be established within the same laboratory setting.

## High spectral-efficiency SDM transmission

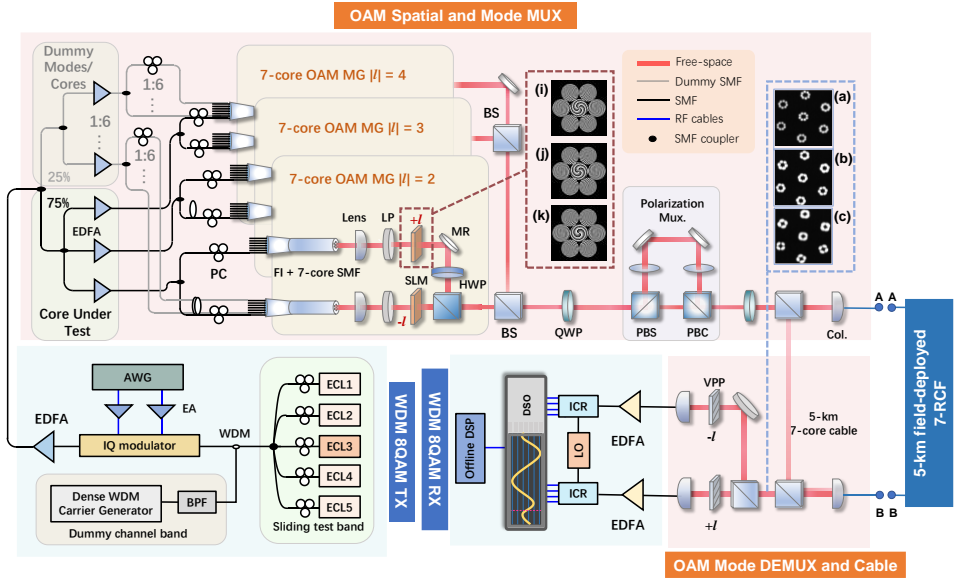

Fig. 3. Experimental setup. ECL: external cavity laser; AWG: arbitrary waveform generator; EA: electrical amplifier; EDFA: erbium-doped fiber amplifier; LP: linear polarizer; SLM: spatial light modulator; MR: mirror; HWP: half-wave plate; BS: beam splitter; QWP: quarter-wave plate; PBS: polarization beam splitter; PBC: polarization beam combiner; Col.: collimator; VPP: vortex phase plate; ICR: integrated coherent receiver. The intensity profiles of OAM MGs $|l|$ = (a) 4, (b) 3, and (c) 2 and the phase masks of OAM MGs $|l|$ = (i) 4, (j) 3, and (k) 2.

The experimental setup of the bidirectional OAM-WDM-SDM data transmission is depicted in Fig. 3. The setup comprises five main parts: a WDM signal transmitter with 8QAM modulation, an OAM spatial and mode multiplexing (MUX) module, the 5-km field-deployed 7-RCF, an OAM mode demultiplexing (DEMUX) module and coherent optical receivers followed by DSP including the offline 4 × 4 MIMO equalization.

*WDM-8QAM signal generation*

At the transmitter, to circumvent laboratory resource limitations, sliding test channels with high optical signal-to-noise ratio (OSNR) and dummy channels with low OSNR are multiplexed together to generate 40 WDM carriers. Five optical carriers with wavelengths in a 0.1nm/12.5 GHz grid from external cavity lasers (ECLs) are employed as sliding test bands. The light source of the dummy wavelength band is generated by a WDM carrier generator based on multiple seed light sources that undergo modulation through a cascaded phase modulator and Mach-Zehnder modulator (MZM)[28]. Following this, the test and dummy bands are combined by a wavelength division multiplexer. Subsequently, the WDM carriers are modulated by a 12-Gbaud 8QAM electrical signal sourced from an arbitrary waveform generator (AWG) via an in-phase/quadrature (I/Q) modulator. This modulation process generates a set of 40-channel WDM signals spanning from 1548.16 nm to 1552.16 nm, maintaining a grid of 0.1 nm/12.5 GHz. The sampling rate of the AWG is set to 120 GSa/s and the data sequence is the pseudo-random binary sequence with a pattern length of $2^{18}-1$. The electrical signals are digitally pre-shaped by a Nyquist filter, specifically a raised cosine filter, with a roll-off factor of 0.01 to align with the 12.5 GHz WDM grid. It is noted that due to device resource constraints, all WDM channels are modulated by a single electrical signal. This configuration may result in limited decorrelation among the WDM channels, potentially leading to an overestimation of system performance due to inter-WDM-channel nonlinear effects[29]. However, the high nonlinear threshold of the 7-RCF would ensure that the inter-WDM-channel nonlinear effect has little impact on the system performance[9]. Due to limited equipment resources, only 40 wavelength channels were used in this experiment. Nonetheless, we do not anticipate a significant drop in performance beyond this spectral band. A total of 312 WDM channels spanning from 1538.19 nm to 1602.10 nm were evaluated in the high-capacity SDM systems using a 7-RCF with similar design in[9], and remarkably consistent performance was sustained across all of the WDM channels.

*OAM Spatial and Mode MUX Module*

After being amplified by a high-power erbium-doped fiber amplifier (EDFA), the generated WDM signals are divided into four branches using an optical power splitter. Three of these are amplified and further divided into six branches to implement mode channels with high OSNR in the fiber core under test, while the remaining branch is amplified by the high-power EDFAs and used to implement low OSNR channels in the dummy cores. The generated test and dummy mode/spatial channels are directed toward three different OAM spatial and modal MUX modules. Within each OAM spatial and mode MUX module, two hexagonally packed 7-core SMFs spliced with a fan-in device are included to convert the in-fiber fundamental modes to free-space Gaussian beams. Then two sets of 7-core Gaussian beams are collimated, linear-polarization filtered and finally images onto spatial light modulators (SLMs). The SLMs are configured with phase masks (brown dashed box in the Fig. 3 and detailed parameters can be found in Supplement S5) to generate 7-core OAM beams with topological charge $+l$ or $-l$. Subsequently, the six groups of 7-core OAM beams generated from the three MUX modules are power-combined and directed through a quarter-wave plate (QWP) to facilitate circular polarization conversion. After propagating through the polarization multiplexing module, which comprises a polarization beam splitter (PBS), an optical de-correlation path with a 4-f configuration and a polarization beam combiner (PBC), four OAM modes $<+l, R>$, $<+l, L>$, $<-l, R>$ and $<-l, L>$ are generated within each MG, where $L$ and $R$ refer to the left- and right-handed circular polarization, respectively. Subsequently, the 84 spatial/mode channels (7 cores × 6 OAM modes × 2 orthogonal circular polarizations) each carrying 40 wavelengths are split into two branches equally via a beam splitter (BS), and they are coupled into the field-deployed 7-RCF from both ends, enabling bidirectional transmission. Here both forward and backward transmissions for each mode and wavelength channel share the same laser source, owing to the limited device resources, so the signal transmitted in both directions is the same. This could introduce coherent Rayleigh noise, manifested as the beating noise between the signal in one transmission direction and Rayleigh backscattering noise in the other direction[30–33]. However, this noise can be substantially mitigated through coherent optical detection (refer to Supplement S6).

*OAM Mode DEMUX Module and Coherent Receiver*

After going through the 5-km field-deployed 7-RCF bidirectional transmission, the intensity distribution of received OAM beams is illustrated in the inset figure (blue dashed box) of Fig 6. Due to the coherent superposition of the four degenerate OAM modes within an MG, a linearly polarized (LP) mode-like intensity distribution[34] is observed in each fiber core. Subsequently, the OAM beams are collimated and split into two branches. After going through a vortex phase plate (VPP) that matches the topological charge of OAM mode being tested, each branch is converted into a Gaussian beam and coupled into an SMF-pigtailed dual-polarization integrated coherent receiver (ICR). Due to limitations in equipment resource at the receiver, only the four OAM beams within the same MG are simultaneously mode converted and then detected by the ICRs in only one direction. Subsequently, the eight electrical signals, generated by two ICRs, are sampled and stored using an eight-channel real-time oscilloscope operating at a sampling rate of 80 GSa/s for offline DSP, which includes timing phase recovery, 4 × 4 MIMO equalization based on the constant modulus algorithm, frequency offset estimation and carrier phase estimation. The measurement is repeated until all the mode/spatial channels in both direction have been measured.

*Power Budget Evaluation of the bidirectional OAM-SDM-WDM system*

The power budget of the bidirectional OAM-SDM-WDM experimental system has been comprehensively evaluated, with the results presented in Table 2. The average power of the WDM signals at the input port of the fan-in device is standardized to 18–19 dBm, equivalent to approximately 2 to 3 dBm per WDM channel. This power level is subject to variations that depend on the OAM MG orders, primarily due to variations in insertion loss attributed to the 7-core OAM MUX module. The total insertion loss within the OAM MUX module includes optical element losses, such as those from the 3 dB beam combiner and the SLM, as well as coupling losses into the field-deployed 7-RCF. The observed coupling loss to the field-deployed 7-RCF for OAM MG $|l| = 4$ is approximately 4 dB, which is 1 dB higher than that of OAM MG $|l| = 2$ and 3. This MG-order dependent loss can be attributed to the radial mismatch between the generated free-space OAM beams and the OAM modes supported within the field-deployed 7-RCF. Additionally, alignment errors in the optical elements used for fiber mode coupling may have contributed to this discrepancy. Such challenges can be addressed through the implementation of precise local phase modulation for OAM modes with varying topological charges ($|l|$) in each fiber core and improved alignment of optical elements. Before being coupled into the field-deployed 7-RCF, the OAM beam is evenly divided into two branches and coupled from both ends of the field-deployed 7-RCF to achieve bidirectional transmission, which causes a power loss of 3dB. After passing through the 5-km field-deployed 7-RCF and the OAM DEMUX module, ∼ −24 dBm of optical power is received at the input port of the optical pre-amplifier located before the ICR, higher than the sensitivity (∼ −37 dBm) of the pre-amplifier.

Table 2. Optical power budget evaluation of the bidirectional OAM-SDM-WDM experiment system.

| OAM mode group | $|l| = 2$ | $|l| = 3$ | $|l| = 4$ |
| --- | --- | --- | --- |
| Average power at Fan-In input | 18dBm | 18dBm | 19dBm |
| Average power per wavelength at Fan-In input | 1.98dBm | 1.98dBm | 2.98dBm |
| Insertion Loss of the OAM Mux. Module (including coupling loss) | 13dB | 13dB | 14dB |
| The loss of being divided into two branches for bidirectional transmission | 3dB | 3dB | 3dB |
| Average fiber loss/core | 1.57dB | 1.58dB | 1.72dB |
| Insertion loss of OAM Demux. Module (including coupling loss) | 9dB | 9dB | 9dB |
| Average received power before Pre-amp. EDFA for each fiber core | -24.59dBm | -24.60dBm | -24.74dBm |

## Results and analysis

### Characteristics of the 7-RCF in the installed cable

The 7-RCF within the fiber cable supports a total of 35 OAM mode groups (MGs), encompassing 7 cores with 5 OAM MGs each. Notably, each MG comprises degenerate OAM modes with topological charge values $+l$ and $-l$ ($l$ = 1, 2, 3 and 4) alongside the OAM MG with a topological charge $l$ = 0. Each of these modes carries two orthogonal polarizations.

The attenuations of all the MGs at the wavelength of 1550 nm are measured by means of optical time-domain reflectometry (OTDR) as detailed in Supplement S2. As depicted in Fig. 4, the attenuation for the OAM MGs of the 7-RCF within the installed cable averages 0.32 dB/km (seeing red triangles in Fig. 4). This is approximately 0.02 dB/km higher than that of the 7-RCF spools (as indicated by the blue circles in Fig. 4). This slight increase can be ascribed to the minor micro-bending and additional stress incurred during the cabling process. Notably, there is a pronounced attenuation difference for the highest-order OAM MG with a topological charge of $|l|$ = 4 in fiber core #1, observed in both the fiber spools and the installed cable. This discrepancy could stem from deformation in the structure of this particular core during the drawing process.

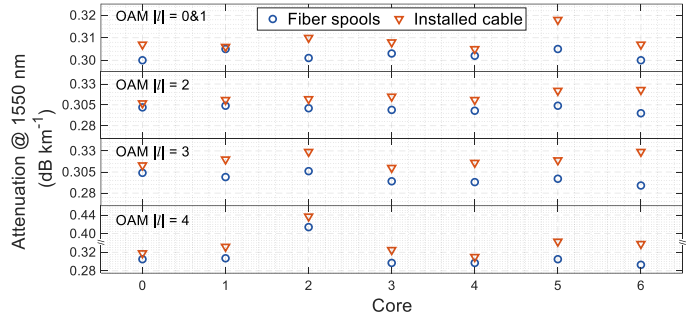

Fig. 4. The measured mode-dependent attenuation of 7-RCF spools and installed cable at 1550 nm.

The inter-MG crosstalk (XT) within an individual fiber core, as well as the inter-core XT of the 7-RCF, have been experimentally characterized for both the installed cable and fiber spools. The measurement details are provided in Supplement S3. As illustrated in Fig. 5, the aggregated crosstalk among OAM MGs with $|l|$ = 2, $|l|$ = 3, and $|l|$ = 4 within the same 7-RCF fiber core of the installed cable is below −12 dB at a wavelength of 1550 nm, while this value is around -20 dB among different fiber cores. In this case, MGs $|l|$ = 0 and $|l|$ = 1 are not used as spatial channels in this work due to their strong inter-MG crosstalk. Therefore, the ring-core design can be further refined to reduce crosstalk between these two MGs, ensuring that all MGs exhibit low inter-MG crosstalk and can be utilized effectively[25]. This performance closely mirrors that of the 7-RCF spool with the length of 14 km or even 34 km. Such consistency arises because the crosstalk is predominantly caused by the OAM mode conversion and core coupling modules at both ends of the 7-RCF (as illustrated in Fig. S3 in Supplement S3), rather than by the fiber itself. We previously consistently measured typical in-fibre XT in the order of -35 dB/km or less in fiber spools[9, 26].

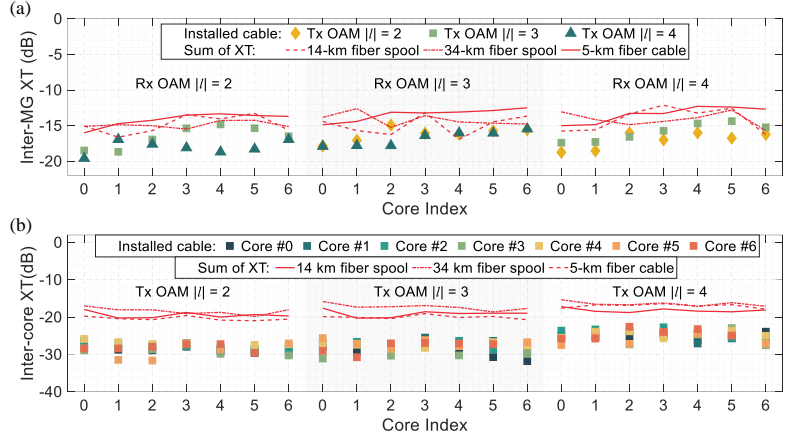

Fig. 5. The measured crosstalk of 7-RCF spools and installed cable at 1550 nm: (a) inter-MG crosstalk within the one single fiber core; (b) inter-core crosstalk.

The differential group delay (DGD) of the OAM MGs are measured with a vector network analyzer (VNA) using a time-domain impulse response method[27], detailed in Supplement S4. As depicted in Fig. 6, large DGD with values of more than 5 ns/km between adjacent pairs of OAM MGs of topological charge $|l| > 1$ can be achieved in each fiber core. This observation indicates weak coupling among high-order OAM MGs within each fiber core[27]. This performance also closely aligns with that of the 7-RCF spool, implying that the effect of cabling and installation on the DGD is negligible.

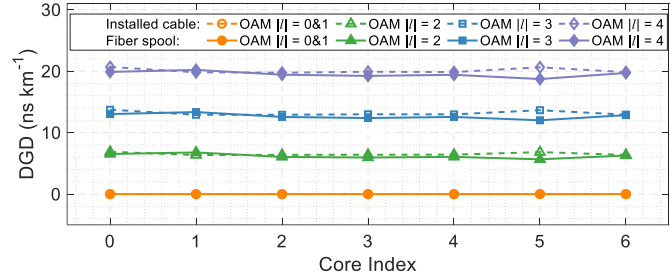

Fig. 6. The measured DGD at 1550nm of 7-RCF spools and installed cable. One averaged DGD value is given for OAM MGs $|l|$ = 0 & 1 in the measured results, as their impulse response merged into one Gaussian-distribution peak due to strong coupling between these two MGs after transmission.

**BER Evaluation for Bidirectional SDM Transmission**

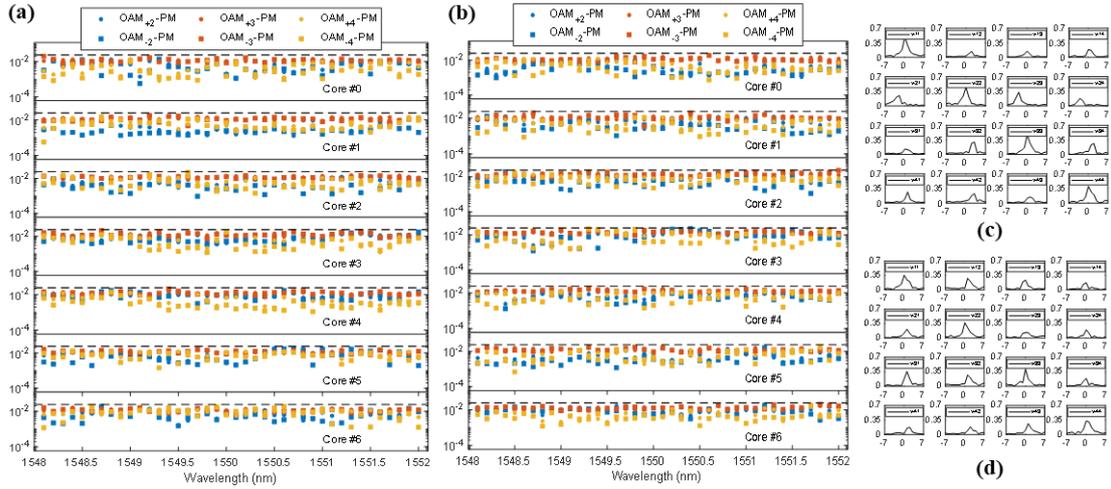

Fig. 7. The measured BERs of all channels after 5-km field-deployed 7-RCF bidirectional transmission: (a) forward transmission and (b) backward transmission; the absolute values of tap weights in 16 FIR filters of 4 ×4 MIMO equalizers to equalize the four modes belonging to OAM MGs $|l|$ = 3 at 1540 nm in the (c) forward transmission and (d) backward transmission.

The measured forward and reverse transmission bit-error rate (BER) values of all 6720 channels (2 direction × 7 cores × 6 OAM modes × 2 polarizations × 40 WDM channels) are shown in Fig. 7(a) and Fig. 7(b). For convenience, the BER evaluation for two orthogonal polarizations is performed together, resulting in the display of only two values within each MG. Notice that the BER performance of the OAM MG with topological charge $|l|$ = 3 is inferior compared to the other OAM MGs, which is primarily attributed to the relatively higher crosstalk from the adjacent two OAM MGs ($|l|$ = 2 and $|l|$ = 4). However, the BER values of all bidirectional channels are below the 20% soft-decision FEC threshold of $2.4\times10^{-2}$, only utilizing modular 4×4 MIMO equalization with a TDE tap number not exceeding 15, as shown in Fig. 7 (c) and (d).

**Effect of Rayleigh Backscattering Noise**

The signal-to-noise ratio (SNR) of the forward transmission signals are experimentally evaluated at varying levels of backward transmission signal power within the same fiber core, to assess the effect of RB noise in the bidirectional transmission system with coherent optical detection. As illustrated in Fig. 8(a), the SNRs of forward transmission signals decrease as backward transmission signal power increases. Fig. 8(b) provides a comparison by illustrating the theoretically calculated signal-to-RB-noise power ratio after coherent detection at different levels of backward transmission power. It is noteworthy that the received signal and RB noise, post-coherent detection, are proportional to the photocurrents generated through the beating process between the optical signal or RB noise and the optical light from the local oscillator (LO) at the coherent optical receiver. As a result, the power ratio between the detected signal and RB noise are expressed as $\langle I_{S\text{-}lo}\rangle^2/\langle I_{RB\text{-}lo}\rangle^2$ in Fig. 8(b). Further details regarding the theoretical RB analysis in the ring-core fiber together with coherent optical detection can be found in Supplement S6.

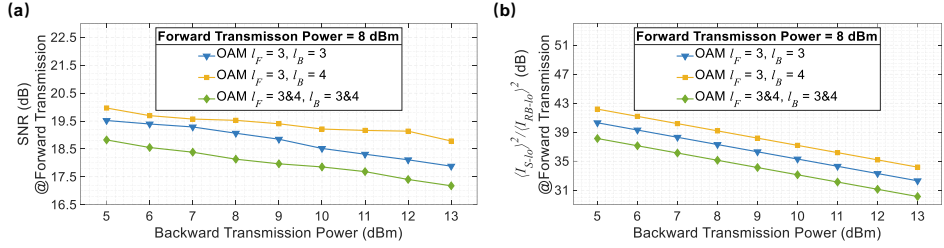

Fig. 8. The measured SNRs under different backward transmission power with 8 dBm forward transmission power after (a) 5-km installed cable transmission and (b) 5-km theoretical transmission. $l_F$: topological charge of the forward transmission OAM light beam; $l_B$: topological charge of the backward transmission OAM light beam; $\langle I_{S-lo}\rangle^2/\langle I_{RB-lo}\rangle^2$ represents the power ratio between the detected signal and RB noise after coherent detection (details can be found in Supplement S6).

An observation from Fig. 8(a) and (b) reveals that the experimentally evaluated signal SNR is higher than the theoretically predicted power ratio between received signals and RB noise. This discrepancy is primarily attributed to various factors beyond RB noise in the experimental system, such as Fresnel reflection of backward injected optical power at the fiber end facet, amplified spontaneous emission (ASE) noise of the EDFAs, inter-mode crosstalk, electrical amplifier noise, quantization noise from the digital-to-analog converter (DAC) in the AWG, and analog-to-digital converter (ADC) within the real-time oscilloscope.

Since the optical carriers of the forward and backward transmissions for each mode and wavelength channel share the same laser source due to equipment resource limitation (seeing Section *OAM Spatial and Mode MUX Module*), beating noise is anticipated to exist between RB noise from the backward transmission and signals transmitted in the forward direction, with its power theoretically being much higher than that of the RB noise itself[35]. However, coherent detection utilizing balanced detectors integrated into the ICR significantly mitigates such signal-RB beating noise, as explained in Supplement S6. In the experimental system, the impact of Fresnel reflections of the backward transmission signals at the fiber end facet, typically characterized by a power considerably higher than that of the RB noise[36, 37], is more pronounced on forward transmission signals at their receiving end. However, noise related to backward transmission, including both Fresnel reflection and RB noise, constitutes a small portion of total noise. Consequently, the experimentally evaluated SNR variation ranges, as depicted in Fig. 8(a) with changing backward transmission signal power, are smaller than those theoretically predicted in Fig. 8(b).

Fig. 8(b) also illustrates that the signal-to-RB-noise power ratio is higher when forward and backward transmission channels share the same mode index ($l_F = l_B$, blue triangle line in Fig. 8) compared to cases with different mode indexes ($l_F \neq l_B$, orange square line in Fig. 8), given fixed forward and backward transmission powers. This is because the mode recapture factor, which represents the proportion of the total scattered power of the backward transmitting mode channel recaptured by the forward transmitting mode channel, is theoretically higher in the former case[38]. Detailed analysis can be found in Supplement S6. Similar trends are observed in the experimentally evaluated results shown in Fig. 8(a). However, we attribute this phenomenon mainly to the further suppression of Fresnel reflection power by the OAM mode DEMUX module when the backward transmission mode index is different from the forward transmission one.

For mode-multiplexed transmission in both forward and backward directions, the theoretical analysis indicates a deterioration in the signal-to-RB-noise power ratio compared to single-mode transmission cases. This is due to the contribution of RB noise from two multiplexed backward transmission modes [e.g., $l_B$ = 3 and 4 in Fig. 8(b)] to the target evaluated forward transmission mode [e.g., $l_F$= 3 in Fig. 8(b)]. In the experimental evaluation, the degradation in SNR performance in mode-multiplexed transmission results not only from increased backward-transmission-power-dependent noise but also from inter-mode crosstalk.

Based on the above analysis, it is found that in short-distance bidirectional transmission systems, the effect of Fresnel reflection at the fiber end facet surpasses that of RB noise. To enhance the SNR of the received signal, one effective strategy involves mitigating Fresnel reflection, such as by applying an anti-reflective film to the fiber end facet. Furthermore, in contrast to SMF based bidirectional transmission systems, the received signal performance is affected not only by RB noise but also by inter-mode crosstalk. In future investigations, we aim to conduct a comprehensive assessment of the collective impact of these two noise sources on the performance of transmission systems at varying distances. This evaluation will contribute to establishing the appropriate transmission distance range for bidirectional SDM transmissions with weakly coupled mode channels.

**Stability of Transmission System over Field-Deployed 7-RCF**

The fiber cable was laid in standard conduits without specialized fastening methods. Portions of these cables hang freely, as shown in Fig. 2(e). Environmental elements such as wind, vehicular traffic, temperature shifts, and humidity variations are unavoidable.

These environmental factors induce random coupling among the degenerate modes within the same mode group of a single fiber core, which leads to fluctuations in the received power over time when only one intra-group mode is received. As a result, measuring the temporal variation of this power fluctuation, whose details can be found in Supplement S7, allows us to gauge the influence of external environmental factors on the intra-group mode coupling. According to the measured results (seeing Fig. S7 in Supplement S7), the power fluctuation of the field-deployed fiber cable due to the random coupling among intra-group modes changes more rapidly than what is seen in optical fiber spools used in laboratory settings, even though the former exhibits relatively smaller magnitude of the power fluctuation due to the protection of the outer sheath in the cable. However, our adaptive MIMO equalization algorithm effectively tracks such variations. In addition, the modal differential delay (DMD) among the intra-group modes remains relatively low, although their values change over time due to random strong coupling. For a transmitting symbol rate of 12GBaud, 15 taps of the time-domain FIR filters for MIMO equalization adequately cover such DMD variations.

In addition, as illustrated in Fig. 5, the measured data show no significant variation in either inter-mode-group or inter-core crosstalk for the field-deployed fiber cable relative to the laboratory fiber spools. This is primarily attributed to the MUX/DEMUX modules, which limit the inter-mode-group and inter-core crosstalk of the entire transmission system. Consequently, variations in the minimal inter-mode-group and inter-core crosstalk of the fiber cable have negligible impact on system performance.

**Conclusions and discussions**

A bidirectional SDM transmission based a 5-km field-deployed 7-RCF with a cladding diameter of 178 μm is experimentally demonstrated in this paper. Implemented over the 7-RCF supporting 84 OAM mode channels each carrying 40 wavelengths, the SDM-WDM system achieves a raw (net) SE of 483.84 (403.2) bit/s/Hz and a capacity of 241.92 (201.6) Tbit/s by utilizing a bidirectional transmission scheme solely applying modular 4×4 MIMO equalization with a TDE tap number not exceeding 15.

Due to constraints in device resources, the demonstration did not extend to showcasing SDM transmission over field-deployed fiber cables at longer distances using the recirculating loop scheme. Nonetheless, the 7-RCF employed in our demonstration exhibits the potential to enable longer-distance SDM transmission with high SE and capacity, whilst keeping low MIMO complexity, and consequently, low cost and low power consumption. We have successfully demonstrated high SE / high capacity SDM transmissions over 7-RCF spools covering distances of 14 km[16], 34 km[9], and 60 km[6], utilizing 4×4 MIMO equalization with low TDE tap numbers, thus ensuring low MIMO complexity. Additionally, we showcased a 100-km transmission supported by a single-core RCF, employing 4×4 MIMO processing with a low TDE tap number to address crosstalk among the degenerate intra-MGs[39].

The good performance of the SDM transmission for up-scaling of SE and capacity per optical fiber while keeping low MIMO complexity within distances of tens of or even one hundred kilometers are realized by exploiting the uniquely excellent characteristics of OAM modes in RCFs. The radial confinement of the ring-shape core of the RCF makes the number of degenerate (or near degenerate) modes in the OAM MGs fixed at four when their topological charge $|l| \geq 1$ (OAM modes with topological charge value $+l$ and $-l$ each carrying two orthogonal polarizations[40], whose differential mode delays are very low due to their good degeneracy). The differential effective refractive index $\Delta n_{eff}$ between adjacent MGs increases with $|l|$[34], which leads to reduced inter-MG coupling therefore good scalability to high-order mode space. Therefore, MIMO processing with a small 4×4 scale and a low TDE tap number and thus low complexity[41–45] can be sustained in the OAM-RCF based systems when more SDM channels are involved to realize high SE/capacity. In our 7-RCF OAM MUX module, mode multiplexing currently relies on a straightforward power combining strategy, leading to insertion losses that increase with the number of MGs utilized. While this method is sufficient for a small set of modes, it becomes less efficient as the number of mode channels grows, with power combining losses becoming more pronounced. Exploring advanced techniques such as multi-core spiral transformation[46] or multi-plane light conversion[47] could offer better power efficiency. Additionally, there's a clear advantage in developing MUX/DEMUX modules that are more integrated. However, such advancements require tackling the significant challenge of converting densely packed input Gaussian beams into a complex 2D input fiber array, a task that presents notable optical engineering challenges.

While the inter-MG crosstalk of the 5-km 7-RCF in field-deployed cable remains lower than that of spatial and mode MUX/DEMUX modules, we anticipate a degree of performance degradation for field-deployed fiber cables compared to lab-deployed optical fiber spools. For instance, fiber fusion errors during the laying process may elevate the inter-MG crosstalk of the RCF[48]. However, judicious control of the fiber fusion error range (e.g., no more than 0.4-μm radial misalignment for the RCF[49]) can effectively manage the increase in inter-MG crosstalk, keeping it to a reasonably low level. By overcoming additional challenges in field-deployed fiber cables for practical application in the future, SDM schemes utilizing OAM modes in RCFs stand as a promising candidate for next-generation high SE/capacity transmission over optical fibers with distances of tens of kilometers (e.g., metro areas, inter-data center links, etc.), where weak coupling among the non-degenerate modes within each fiber core is achievable, and in-line optical amplification towards FM-MCFs is unnecessary.

## Acknowledgements

The work was supported by National Natural Science Foundation of China (62335019, 62101602) and National Key R&D Program of China (2018YFB1801800).

## Authors' contributions

Junyi Liu, Zengquan Xu and Jie Liu conceived the original idea and designed the experiment. Junyi Liu, Zengquan Xu, Yuming Huang and Yining Huang conducted the experiment. Shuqi Mo analyzed the effect of Rayleigh backscattering noise. Zhenhua Li and Yuying Guo experimentally realized the generation of wideband WDM carriers. All authors analyzed their experimental results and contributed to writing and proofreading the manuscript. Lei Shen and Shuo Xu contributed to the cable fabrication. Cheng Du, Qian Feng and Jie Luo deployed the cable. Siyuan Yu and Jie Liu supervised the overall project.

## Authors' information

Not applicable.

## Conflict of Interest

The authors declare no conflict of interest.

## Availability of data and materials

The experimental data that support the works of this study are available from the corresponding authors on reasonable request.

**Supplementary information** is available for this paper

## Declarations

## Ethical Approval and Consent to participate

There is no ethics issue for this paper.

## Consent for publication

All authors agreed to publish this paper.

## competing interests

The authors declare no competing fnancial interests